\documentclass[conference]{IEEEtran}
\IEEEoverridecommandlockouts
\usepackage{cite}
\usepackage{amsmath,amssymb,amsfonts}
\usepackage{algorithmic}
\usepackage{graphicx}
\usepackage{textcomp}
\usepackage{xcolor}
\usepackage{subcaption}
\usepackage{wrapfig}
\def\BibTeX{{\rm B\kern-.05em{\sc i\kern-.025em b}\kern-.08em
    T\kern-.1667em\lower.7ex\hbox{E}\kern-.125emX}}
\begin{document}

\title{Volume Tracking Based Reference Mesh Extraction for Time-Varying Mesh Compression\\

}

\author{\IEEEauthorblockN{Guodong Chen\IEEEauthorrefmark{1},
        Libor Váša\IEEEauthorrefmark{2},
        Fulin Wang\IEEEauthorrefmark{3},
        Mallesham Dasari\IEEEauthorrefmark{1}}
        \IEEEauthorblockA{\IEEEauthorrefmark{1}Northeastern University,
 			\IEEEauthorrefmark{2}University of West Bohemia,
                \IEEEauthorrefmark{3}Tsinghua University\\
 			\IEEEauthorrefmark{1}\{chen.guod, m.dasari\}@northeastern.edu
                \IEEEauthorrefmark{2}lvasa@kiv.zcu.cz
 			\IEEEauthorrefmark{3}wfl22@mails.tsinghua.edu.cn
 			}}


\maketitle

\begin{abstract}
Time-Varying meshes (TVMs), characterized by their varying connectivity and number of vertices, hold significant potential in immersive media and other various applications. However, their practical utilization is challenging due to their time-varying features and large file sizes. Creating a reference mesh that contains the most essential information is a promising approach to utilizing shared information within TVMs to reduce storage and transmission costs. We propose a novel method that employs volume tracking to extract reference meshes. First, we adopt as-rigid-as-possible (ARAP) volume tracking on TVMs to get the volume centers for each mesh. Then, we use multidimensional scaling (MDS) to get reference centers that ensure the reference mesh avoids self-contact regions. Finally, we map the vertices of the meshes to reference centers and extract the reference mesh. Our approach offers a feasible solution for extracting reference meshes that can serve multiple purposes such as establishing surface correspondence, deforming the reference mesh to different shapes for I-frame based mesh compression, or defining the global shape of the TVMs.

\end{abstract}


\section{Introduction}
With the rapid advancement of 3D sensing technologies, acquiring complex 3D content with high levels of detail and quality has become increasingly feasible. 3D meshes are commonly used to represent complex 3D content, especially in Augmented Reality (AR) and Virtual Reality (VR) environments. Unlike 3D animation, these 3D meshes are Time-Varying meshes (TVMs) that consist of a set of mesh frames that may include time-varying vertex positions, face connectivity, and attributes. Therefore, they require substantial data for storage and transmission, and processing these 3D meshes poses additional challenges.

In recent years, an increasing number of methods have attempted to utilize spatial and temporal correlation to encode or decode meshes more efficiently. Existing methods focus on synthesized mesh sequences with consistent topology across different frames, leveraging temporal correlation to improve compression. However, these approaches require the meshes to maintain the same topology, similar to 3D animation, which limits their practicality in real-world applications.

The Moving Picture Experts Group (MPEG) has developed several mesh processing standards, including IC, MESHGRID, and FAMC \cite{FAMC2021}. However, these standards are limited to processing dynamic mesh sequences with constant connectivity and cannot handle TVMs with changing topology, geometry, and attribute information. To address this issue, the MPEG 3D Graphics Coding (3DG) group recently issued a Call for Proposals (CfP) \cite{cfp2021} for the Video-based Dynamic Mesh Coding (V-DMC) standard. In response, Apple proposed a video and subdivision-based mesh coding (VSMC) scheme \cite{apple2022VSMC}, which was then selected as the foundation for the V-DMC standard. However, the temporally consistent re-meshing process required by VSMC is not always feasible, which imposes constraints on TVMs. Thus, efficient and effective 3D mesh processing remains an open challenge.

Recently, embedded deformation has been used in 3D mesh processing \cite{Embedded Deformation KDDI, Embedded Deformation, Embedded Deformation UCSD}. TVMs lack explicit vertex correspondence between frames and usually have different numbers of vertices and varying connectivity over time, so representing inter-frame differences is challenging. Embedded deformation addresses this problem by transforming the reference frame into a similar shape to the current frame while preserving its topology. To improve results, these methods often divide TVMs into several groups and select a frame as the reference mesh for further deformation. However, this approach faces limitations when self-contact occurs, or new objects emerge in the reference mesh, often resulting in visual artifacts or inaccuracies in the mesh representation. Without a robust method for creating a self-contact-free reference mesh, the potential of embedded and other deformation methods cannot be fully realized.

In this paper, we propose a novel volume tracking-based reference mesh extraction method that utilizes the volume inside meshes. Initially, ARAP volume tracking identifies volume centers within each mesh frame. These centers are processed using MDS to derive reference centers. We map the vertices of each frame to these reference centers using RBFs, optimizing the alignment to produce a well-aligned reference mesh. Finally, Poisson Surface Reconstruction extracts the reference mesh that contains every separate component without self-contact. The main contributions of this paper can be summarized as follows:
\begin{itemize}
    \item We propose a novel reference mesh extraction method for TVMs that leverages volume tracking to create a reference mesh without self-contact regions.
    \item We present an optimization technique designed to align a group of meshes with different topologies, ensuring the extraction of one reference mesh.
    \item We design an improved vertex registration method that can accurately establish correspondence of vertices lying in similar positions in the objects in different frames.
\end{itemize}


\section{Related work}\label{Related work}
Our method focuses on extracting a reference mesh from a group of consecutive mesh frames that can serve as a foundation for various applications, such as surface registration and I-frame-based mesh compression. In this section, we provide a detailed overview of some existing mesh compression methods that could benefit from the reference mesh. 

\subsection{Dynamic mesh processing}
In 2021, the MPEG 3DG group issued a Call for Proposals on dynamic mesh coding, aiming to develop a new standard under the Visual Volumetric Video-Based Coding (V3C) standard, known as the V-DMC standard~\cite{cfp2021}. Then the VSMC scheme proposed by Apple was selected as the basis for the V-DMC standard~\cite{apple2022VSMC}. V-DMC uses inter-frame coding for dynamic meshes, representing a mesh with a decimated base mesh and a set of displacements. It estimates a motion field by tracking corresponding vertices between the current and reference meshes. 
However, to leverage temporal correlation, V-DMC requires the input mesh sequence to have a constant topology or use time-consistent re-meshing. Since VSMC's time-consistent re-meshing is not always feasible, V-DMC cannot perform well on TVMs. 

\subsection{Time-varying mesh processing}
Modern 3D reconstruction techniques use a combination of RGB cameras, infrared (IR) cameras, and unstructured static IR laser light sources to generate detailed 3D meshes. These meshes are typically time-varying, and processing and utilizing them is a major focus of research in this field.

Early works such as \cite{Progressive Coding of 3D Dynamic Mesh Sequences Using Spatiotemporal Decomposition} used the first mesh as the 'reference mesh'. This first mesh is converted into a semi-regular normal mesh, followed by motion estimation to map it onto the following meshes. However, this method often brings significant artifacts and distortions, especially when dealing with a large number of meshes. Additionally, if the first mesh contains self-contact parts that tear apart in subsequent meshes, it can result in later meshes reconstructed from non-existent parts, further increasing the distortion.

In 2015, Collet et al. \cite{High-quality streamable free-viewpoint video} estimated a feasibility score of each frame being a keyframe. They selected a relatively small set of meshes as keyframes and enforced that each keyframe covers a continuous frame range. By doing this they could finally find the minimum set of keyframes that can register all other frames within some tolerance. Also, they considered the self-contact problem. In their keyframe choosing algorithm, they used larger area, lower-genus surface, and more connected components as the standard of identifying keyframe, which solved the self-contact problem to a certain extent. However, this method will repeat this process on the remaining frames until every frame is associated with a keyframe, which is computationally expensive. Complex self-contact scenarios might still pose challenges, potentially affecting the fidelity of the keyframe representation.

\subsection{Embedded deformation processing}
\begin{wrapfigure}{r}{0.2\textwidth}
  \vspace{-0.3in}
  \begin{center}
    \includegraphics[width=\linewidth]{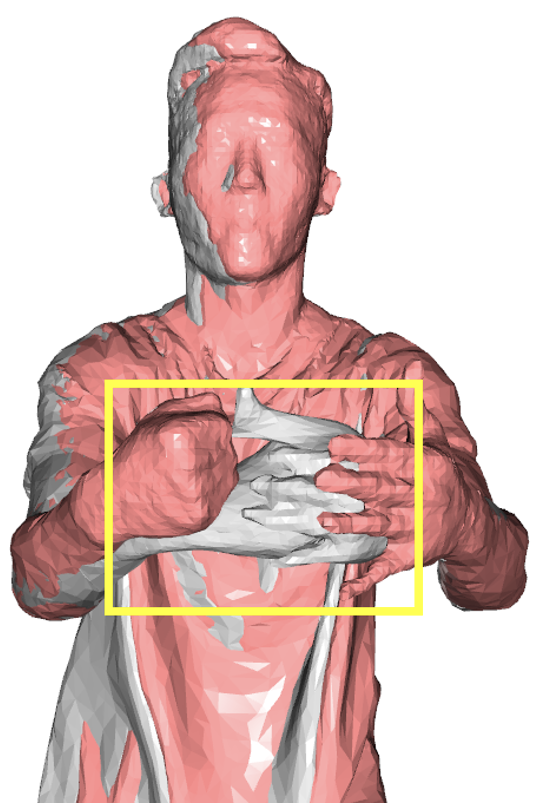}
  \end{center}
  \vspace{-0.15in}
  \caption{Distortions caused by self-contact region.} 
  \vspace{-0.12in}
  \label{self_contact_distortions}
\end{wrapfigure}
Embedded deformation is a technique used in computer graphics and computational geometry to enable smooth, continuous deformations of complex 3D models. For 3D mesh deformation, certain vertices are strategically selected to drive the deformation process. Each selected node influences the surrounding vertices, allowing localized transformations. Based on the movements of these selected nodes, a geometric transformation operator is applied to each vertex in the previous mesh to 'predict' the following meshes, whether they are dynamic meshes or TVMs. Instead of saving all information for each mesh, embedded deformation allows saving a group of meshes with one reference mesh and several underlying transformations~\cite{Embedded Deformation UCSD}\cite{Embedded Deformation KDDI}.
However, these deformation methods apply transformations to every vertex in the reference mesh, which can cause significant distortions when the reference mesh contains self-contact regions, as illustrated in Fig. \ref{self_contact_distortions}. Additionally, new objects appearing in the processing group can also cause distortions. To deform meshes correctly, we have to find an additional approach to handle self-contact regions or divide meshes with self-contact regions into different groups. However, identifying these self-contact regions is challenging.

\section{Proposed method} \label{Proposed method}

Observing these challenges, we propose a volume tracking-based reference mesh extraction method that could create a self-contact-free reference mesh. Our method uses volume tracking to address the self-contact problem robustly, a persistent issue in the mesh compression domain that few existing approaches adequately tackle. With this valuable reference mesh, we can achieve more efficient and reliable 3D mesh processing such as using as-rigid-as-possible deformation \cite{As-rigid-as-possible deformation} to deform the reference mesh into different shapes for mesh reconstruction. This method focuses on the pre-processing step in 3D mesh processing, which can easily serve as the basis of many mesh compression methods that require reference mesh or need to perform deformation.

\begin{figure}[t]
\centerline{\includegraphics[width=0.5\textwidth]{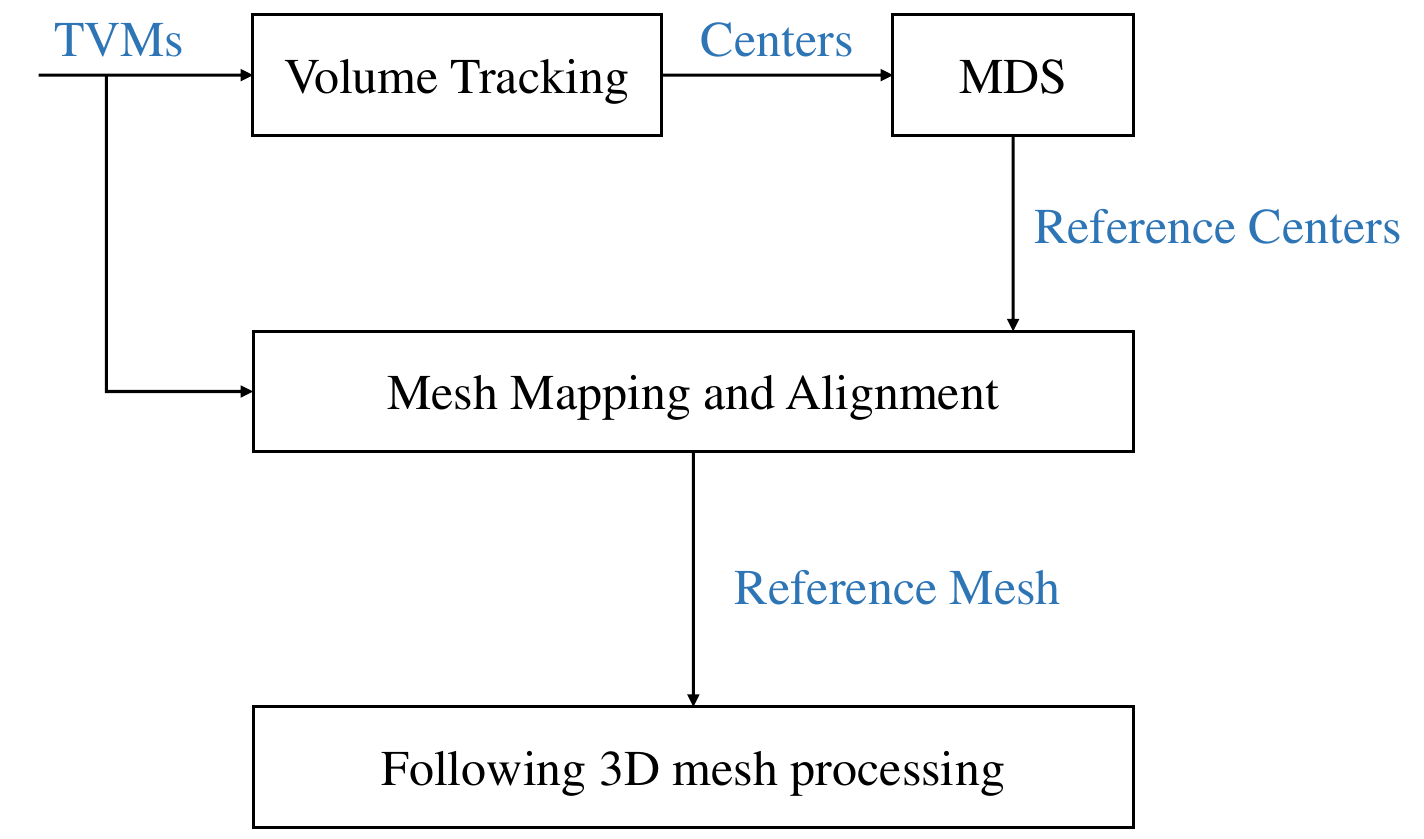}}
\caption{The simplified workflow of our proposed volume tracking-based reference mesh extraction method.}
\label{workflow}
\vspace{-0.25in}
\end{figure}

Fig. \ref{workflow} shows the workflow of our proposed method. The process begins with As-Rigid-As-Possible (ARAP) volume tracking~\cite{MaxAffinityGlobalTracking}, which uniformly distributes a set of centers within the enclosed volume of each mesh frame. These centers are then processed using Multidimensional Scaling (MDS)~\cite{MDS} to derive reference centers that serve as the basis for subsequent operations. We divide the sequence of TVMs into Groups of Frames (GoF) and use Radial Basis Functions (RBFs)~\cite{RBF} to map the vertices of each frame to the reference centers. An optimization step follows to ensure the tight alignment. The resulting aligned vertices are used to generate a reference mesh through Poisson Surface Reconstruction~\cite{Poisson}. This reference mesh is mapped to each mesh in the group using our improved point matching method. Then, we employ as-rigid-as-possible deformation~\cite{As-rigid-as-possible deformation} to generate a re-meshed version of each frame, maintaining the same topology and connectivity as the reference mesh. 

\subsection{As-Rigid-As-Possible Volume Tracking}\label{ARAP volume tracking}
Consider the TVMs sequence, $ S = \{M_0, M_1, \ldots, M_{N-1}\} $, where \( N \) is the total number of frames in the sequence. Let \(M(t) = (V_t, E_t, F_t)\) be a static mesh at time \( t \). \( V_t \), \( E_t \), and \( F_t \) represent the sets of vertices, edges, and faces (triangles) respectively.

We applied the As-Rigid-As-Possible Volume Tracking method, as proposed in \cite{MaxAffinityGlobalTracking}, to the TVMs sequence S. The method identifies a fixed set \( C \) of \( K \) points, referred to as centers, each representing a small volume surrounding it with positions that vary over time. Each center follows a certain trajectory \( c_i = [c_i(0), c_i(1), \ldots, c_i(N-1)] \in \mathbb{R}^{N \times 3} \), where \( N \) is the number of frames and \( c_i(N) \) is the position of the \( i \)-th center in the \( f \)-th frame. The method aims to uniformly distribute the centers inside the enclosed volume of each frame while ensuring that each center moves coherently with its neighboring centers.
For any frame $f$, we define $C_f = \{c_1(f), c_2(f), \ldots, c_K(f)\}$ as the set of all centers' positions in that frame.

\subsection{Extracting Reference Centers: Multidimensional Scaling}
We first build a distance matrix from the largest distances of centers.
The largest distance between two centers \( c_i \) and \( c_j \) is calculated using the Euclidean distance:

\begin{equation}
d_{ij} =\max_{f} \|c_i(f) - c_j(f)\|
\end{equation}

The distance matrix \( D \) is then constructed as:

\begin{equation}
D =  \left( \begin{array}{cccc}
d_{11} & d_{12} & \cdots & d_{1N} \\
d_{21} & d_{22} & \cdots & d_{2N} \\
\vdots & \vdots & \ddots & \vdots \\
d_{N1} & d_{N2} & \cdots & d_{NN}
\end{array} \right)
\end{equation}

Once the distance matrix \( D \) is computed, it is used as the input for Multidimensional Scaling (MDS). MDS seeks to find a set of points \( x_1, x_2, \ldots, x_K \) in a 3-dimensional space that minimizes the stress function given by:

\begin{equation}
\min \quad \sigma(X) = \sqrt{\frac{\sum_{i < j} \left(d_{ij} - \|x_i - x_j\|\right)^2}{\sum_{i < j} d_{ij}^2}}
\end{equation}

Here, \( X = [x_1, x_2, \ldots, x_K] \) represents the reference centers we want in the 3-dimensional space, \( d_{ij} \) are the distances from the matrix \( D_f \), and \( \|x_i - x_j\| \) is the Euclidean distance between points \( x_i \) and \( x_j \) in the reference centers $X$. In the set of reference centers $X$, connected centers should remain contiguous, while those that are in fact disconnected should be apart, since we enforce the maximum distance over the length of the sequence. Finally, we perform a rigid alignment of two sets of 3D points using Singular Value Decomposition (SVD) and call the space where the reference centers live the reference space.

\subsection{Mesh mapping and alignment optimization} \label{reference mesh}
To get better results, we divide the TVMs sequence \( S \) into separate GoF, with each group having its own set of reference centers \( X_g \), where \( g \) indicates the group index. This division allows tailored processing for different segments of the sequence, enhancing the fidelity of the mesh mapping process.

For each group \( g \), and each frame \( f \) within this group, the vertices \( V_f \) of the original mesh are mapped to the reference centers using RBFs. The transformation is defined as follows:

\begin{equation}
V_f' = \text{RBF}(C_{f,g}, X_g)(V_f)
\end{equation}
where \( V_f' \) represents the mapped vertex positions in frame \( f \), \( C_{f,g} \) denote the specific centers used for frame \( f \) within group \( g \), and \( X_g \) denotes the reference centers obtained through MDS for group \( g \).This mapping aims to get a dense set of vertices that serves as the basis for getting a reference mesh.

Due to the inherent diversity in poses and topologies of the meshes within each GoF, the resulting mapped meshes exhibit considerable variation in shape. This lack of cohesiveness among the mapped meshes presents significant challenges for subsequent steps, particularly when attempting to get a single reference mesh that accurately represents the entire group.

In the following, we introduce an optimization-based method for computing the optimal  \( X_{f,g}' \) for each mesh in each GoF to ensure the highest alignment of the mapped meshes. We choose one of the mapped meshes in group \( g \) as the R-frame, with \( V_R \) representing the R-frame's vertex positions. The \( X_{f,g}' \) is determined by solving the following optimization problem:
\begin{equation}
\min_{X_{f,g}'} \;  E_{\text{IoU}}({V_{f}'}, {V_{R}}) 
\end{equation}
where $E_{\text{IoU}}$ measures the Intersection Over Union (IoU) of the optimized mapped mesh and the target reference mesh. $E_{\text{IoU}}$ can be computed by creating voxel grids for meshes and volume sampling.
Specifically, the IoU between two voxel grids \( \mathcal{VG}_1 \) and \( \mathcal{VG}_2 \) can be computed as:

\begin{equation}
\text{IoU} = \frac{|\mathcal{V}_1 \cap \mathcal{V}_2|}{|\mathcal{V}_1 \cup \mathcal{V}_2|}
\end{equation}
where \( \mathcal{V}_1 \) and \( \mathcal{V}_2 \) are the sets of voxel indices for \( \mathcal{VG}_1 \) and \( \mathcal{VG}_2 \), respectively.

Finally, we adopt Poisson Surface Reconstruction \cite{Poisson} on the dense vertices set we've got from above and get the reference mesh \( {M}_{ref} \), which has no self-contact part while containing all components in the group.

\subsection{As-rigid-as-possible deformation re-meshing}\label{ARAPdeformation}
After getting the reference mesh \( {M}_{ref} \) without self-contact regions from Section \ref{reference mesh}, we can process a sequence of meshes more efficiently, e.g. 3D mesh deformation. Next, we describe how to deform this reference mesh to different poses and its potential in the mesh compression.

The ARAP deformation method is a widely used technique in computer graphics for mesh deformation. This method aims to preserve the local geometric details of the mesh by minimizing distortions during the deformation process. ARAP achieves this by enforcing constraints that maintain the local rigidity of the mesh elements, ensuring that transformations applied to the mesh approximate rigid body transformations as closely as possible. This approach is particularly effective in scenarios where the mesh undergoes complex deformations, as it provides a balance between flexibility and structural integrity. Using ARAP deformation, we can deform \( {M}_{ref} \) to \( {M}_{i} \) by controlling a small number of key points.

Specifically, we improved the key point matching approach mentioned in \cite{keypoint matching} to get better matched key points for subsequent ARAP deformation. We first adopt the intrinsic shape signatures (ISS) key points method \cite{ISS} to generate key points for \( {M}_{ref} \) and each mesh in the TVMs sequence S. We denote the set of key points of \( {M}_{ref} \) as \( \{r_1, \ldots, r_m\} \), and the set of key points of \( M(i) \) as \( \{i_1, \ldots, i_n\} \). For each key point \( r_l \) in \( \{r_1, \ldots, r_m\} \), we find a matched point \( i_t \) in \( \{i_1, \ldots, i_n\} \) with the smallest matching error, defined as:
\begin{equation}
\begin{aligned}
\arg \min_{i_t} E_r(i_t) &= \min_{R} \text{Err}(\mathcal{N}(i_t), R(\mathcal{N}(r_l))) \\
& \text{subject to} \quad r_l = i_t \label{find key points}
\end{aligned}
\end{equation}
where \( \mathcal{N}(i_t) \) is the set of neighboring vertices within a radius around \( i_t \) in \( M(i) \), \( \mathcal{N}(r_l) \) is the set of neighboring vertices within a radius around \( r_l \) in \( {M}_{ref} \), and \( R \) is a rotation vector \( \mathbf{r} \in \mathbb{R}^3 \) containing 3 angles that apply rotation to each vertex in \( \mathcal{N}(r_l) \). We constrain \( r_l = i_t \) and ensure that the rotation matrix \( R \) only rotates around the fixed point \( r_l \). The function \(\text{Err}(\cdot)\) is defined as the sum of the average distance between the two vertex sets, computing the distance from the source vertex set to the closest point in the target vertex set and the Hausdorff distance \cite{aspert2002Hausdorff} between the two vertex sets. Additionally, we remove the matched keypoint pairs with large matching errors by verifying whether \( E_r(i_t) \) exceeds a threshold \( \sigma_{\text{th}} \).

\begin{figure}[htbp]
    \centering
    \begin{subfigure}[b]{0.23\textwidth}
        \centering
        \includegraphics[width=\textwidth]{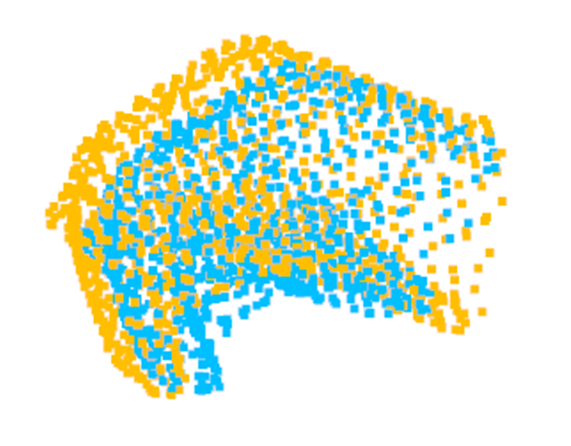}
        \caption{point matching process in \cite{Embedded Deformation KDDI}}
        \label{matching process a}
    \end{subfigure}
    \hfill
    \begin{subfigure}[b]{0.22\textwidth}
        \centering
        \includegraphics[width=\textwidth]{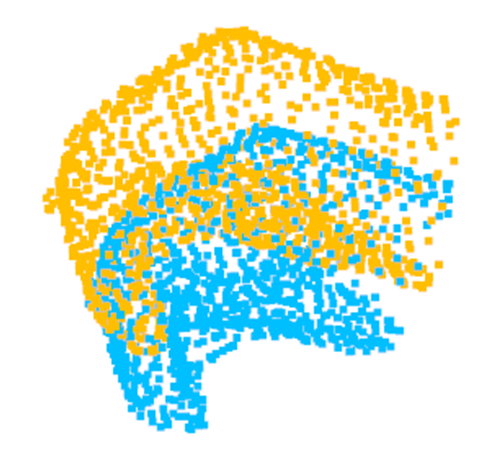}
        \caption{point matching process in our method}
        \label{matching process b}
    \end{subfigure}
    \caption{Visual comparison of key points matching process. The yellow points set represents the key point on the palm and its surrounding neighbors, while the blue points set represents the key point on the back of the hand and its surrounding neighbors. Although these two key points have similar surrounding neighbors, they do not match. In (a), the two key points are mismatched, whereas in (b), our approach successfully identifies their differences.}
    \label{Comparison of matching process}
\end{figure}

\begin{figure}[htbp]
    \centering
    \begin{subfigure}[b]{0.24\textwidth}
        \centering
        \includegraphics[width=\textwidth]{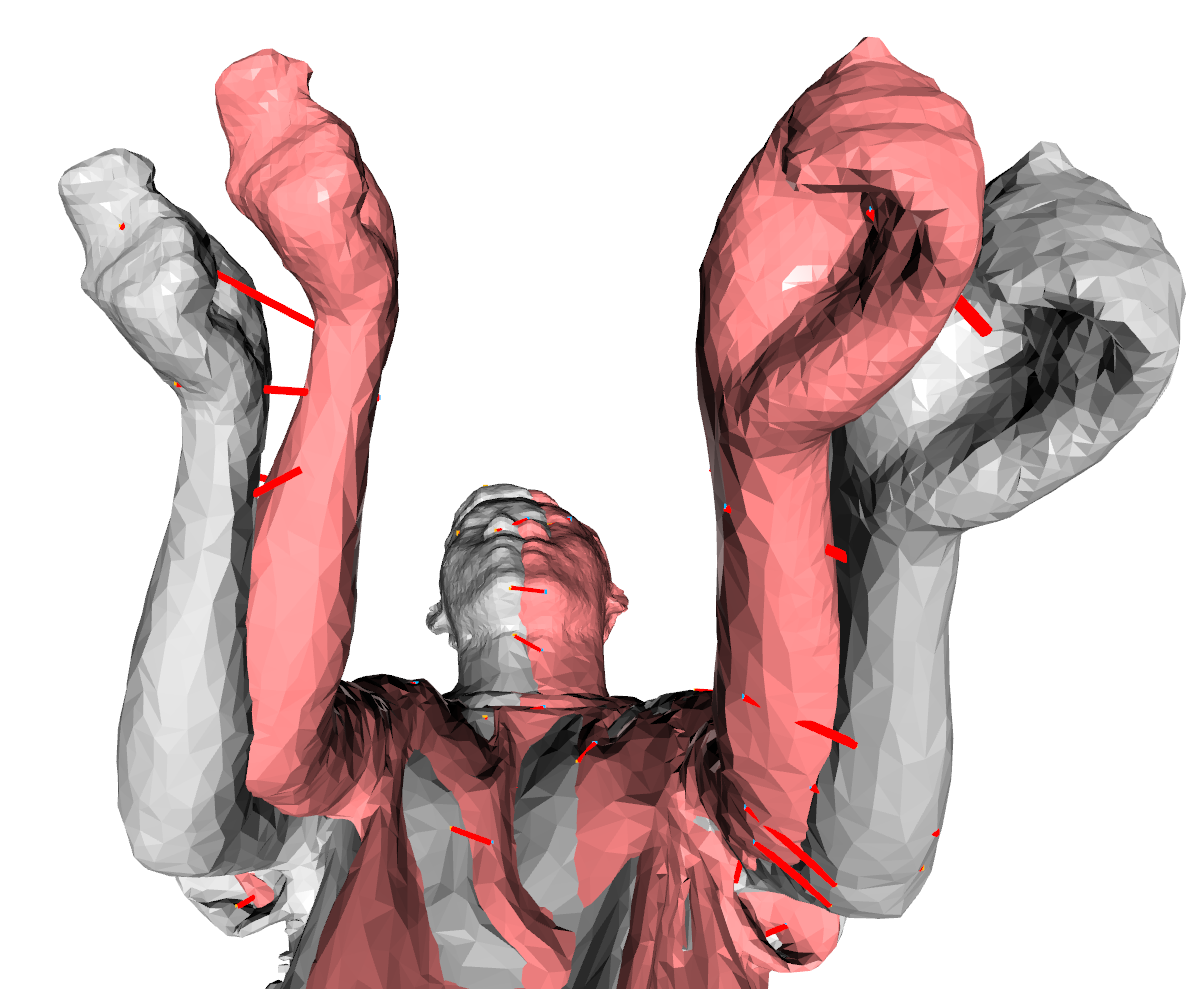}
        \caption{point matching method in \cite{keypoint matching}}
        \label{matched key points a}
    \end{subfigure}
    \hfill
    \begin{subfigure}[b]{0.24\textwidth}
        \centering
        \includegraphics[width=\textwidth]{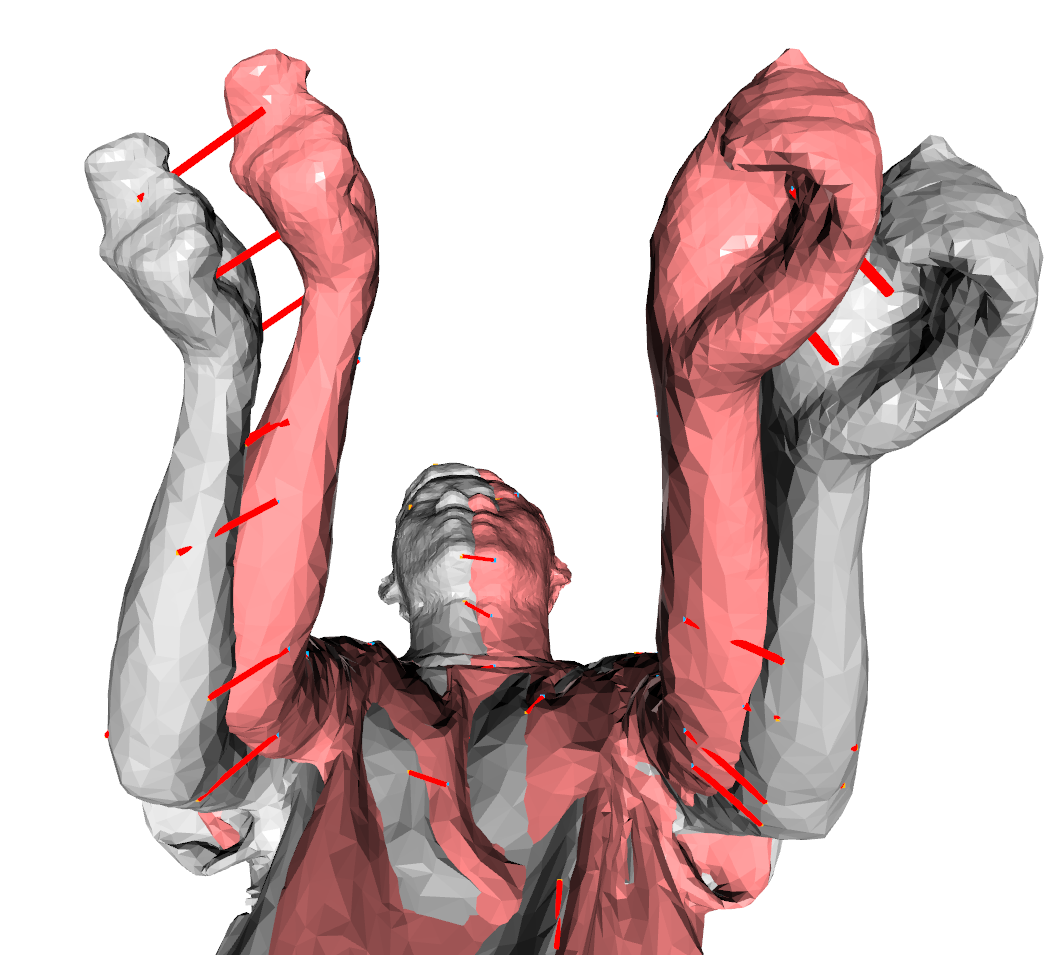}
        \caption{Our point matching method}
        \label{matched key points b}
    \end{subfigure}
    \caption{Visual comparison of matched key points. (a) displays the point matching results from \cite{keypoint matching}, while (b) presents the matching results using our method. The matched key point pairs are connected by red lines. Some connections in (a) on the arm are mismatched, e.g. the key point on the right hand was incorrectly matched to the arm. Compared to (a), the connections in (b) are more accurately matched.}
    \label{Comparison of matching results}
\end{figure}

Fig. \ref{Comparison of matching process} illustrates the difference between our improved method and the original method in judging whether the key point on the palm and the key point on the back of the hand are matched. Before the improvement, the optimization process in \eqref{find key points} would incorrectly judge \( \mathcal{N}(r_l) \) and \( \mathcal{N}(i_t) \) as similar regions, resulting in these two points being mistakenly identified as matched key points, as shown in Fig. \ref{matching process a}. In contrast, our improved method, as shown in Fig. \ref{matching process b} can accurately distinguish that these two points cannot be identified as matched key points though they have similar neighbors. This improvement allows us to achieve better results in finding matched key points, as demonstrated in Fig. \ref{Comparison of matching results}.

\begin{figure*}[tbh]
    \centering
    \begin{subfigure}[b]{0.3\textwidth}
        \centering
        \includegraphics[width=\textwidth]{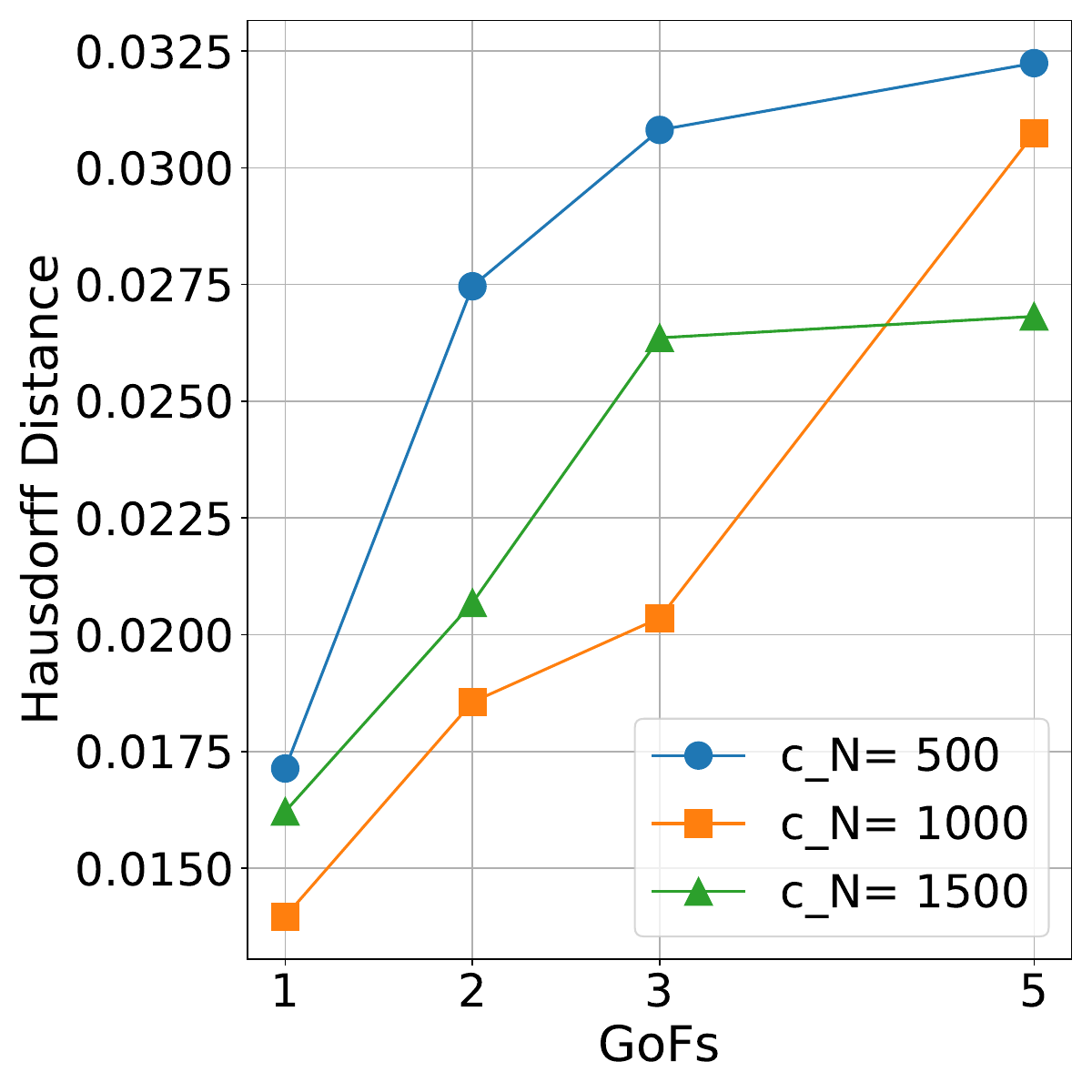}
        \caption{Levi}
        \label{Levi}
    \end{subfigure}
    \hfill
    \begin{subfigure}[b]{0.3\textwidth}
        \centering
        \includegraphics[width=\textwidth]{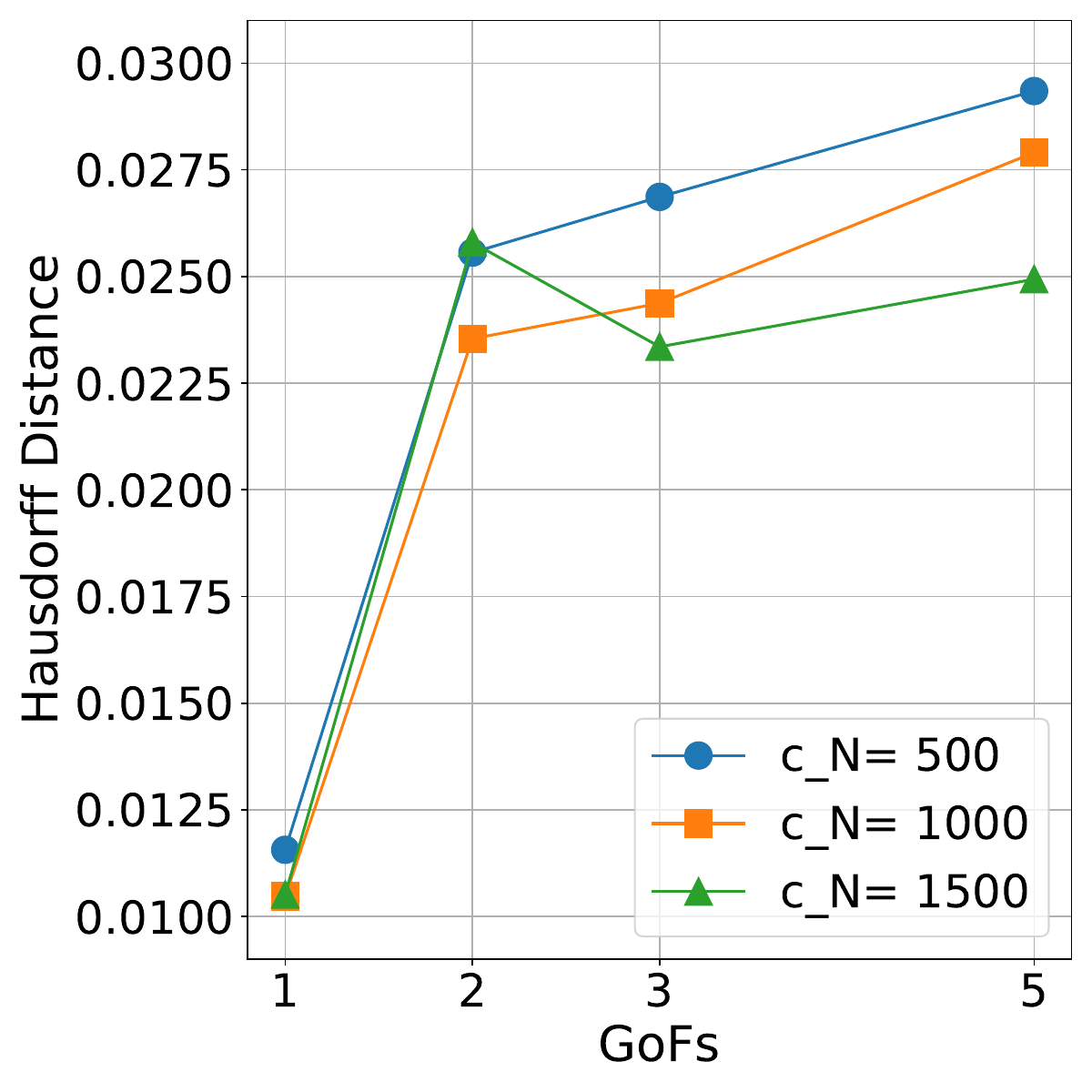}
        \caption{Dancer}
        \label{Dancer}
    \end{subfigure}
    \hfill
    \begin{subfigure}[b]{0.3\textwidth}
        \centering
        \includegraphics[width=\textwidth]{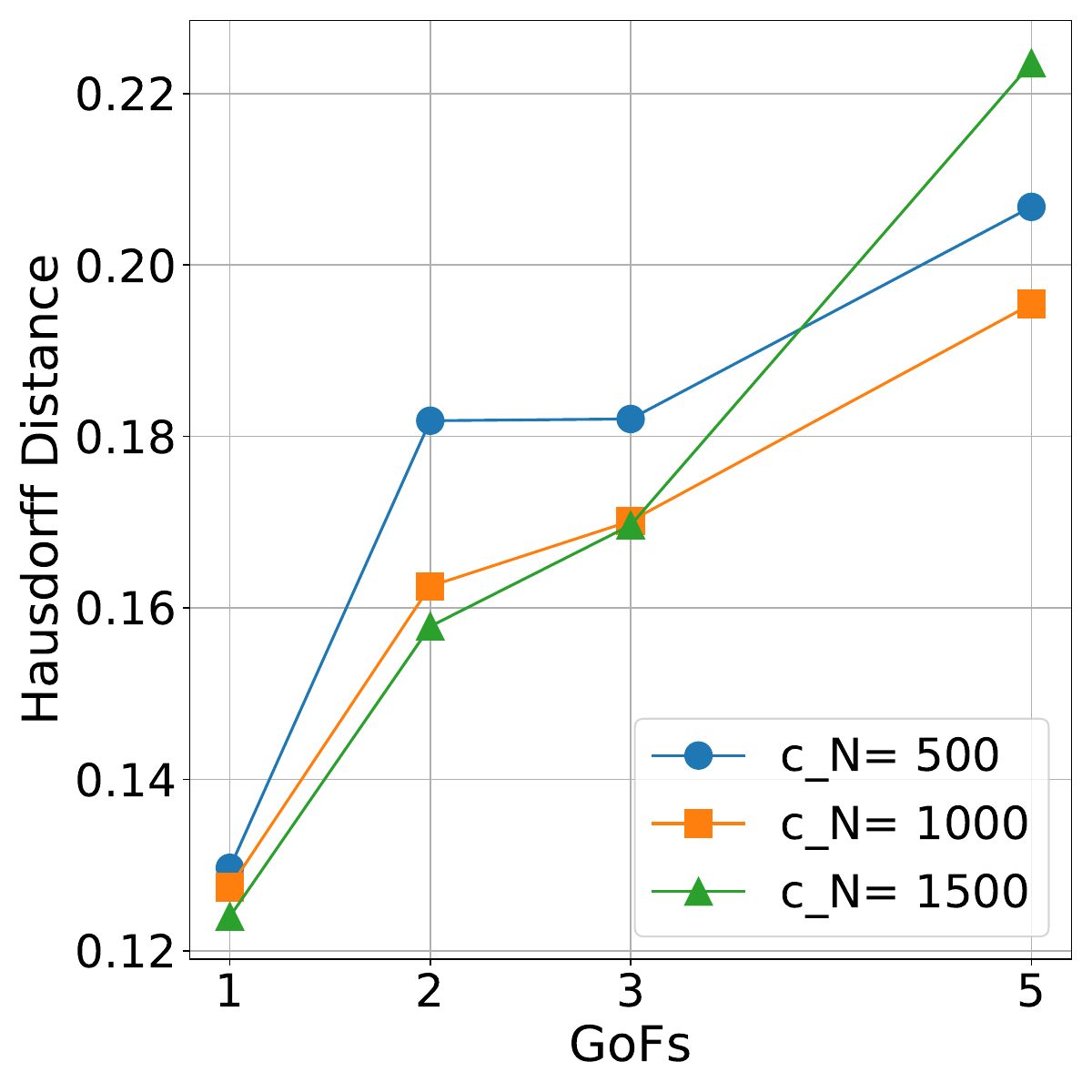}
        \caption{Basketball player}
        \label{Basketball player}
    \end{subfigure}
    \caption{Comparison of Hausdorff Distance for different center numbers across various GoFs. The plots illustrate the performance of three TVMs: (a) Levi, (b) Dancer, and (c) Basketball player. Each graph shows the Hausdorff Distance for three different center numbers: 500 (blue circle), 1000 (orange square), and 1500 (green triangle). The Hausdorff distance measures the geometry distance between vertices, making it sensitive to the scale of mesh sequences. Hence, different mesh sequences will have varying ranges of Hausdorff distances.}
    \label{Hausdorff distance evaluation}
\end{figure*}

After getting well-matched key point pairs, we can use the ARAP deformation method to handle the challenging task of accurately deforming \( {M}_{ref} \) by optimizing the following energy function:
\begin{equation}
\begin{aligned}
\sum^{\left| P_i \right|}_{i} \sum_{P_j \in \mathcal{N}(P_i)} &w_{ij} \left\| ( \mathbf{P}'_i - \mathbf{P}'_j ) - \mathbf{R}_i ( \mathbf{P}_i - \mathbf{P}_j ) \right\|^2 \\
&+ \alpha A \| \mathbf{R}_i - \mathbf{R}_j \|^2
\end{aligned}
\end{equation}
where $\mathcal{N}(P_i)$ represents the set of connected neighbors of vertex $P_i$, $w_{ij}$ are the weights, $\mathbf{P}_i$ and $\mathbf{P}_j$ are the original positions of vertices $i$ and $j$ in \( {M}_{ref} \), $\mathbf{P}'_i$ and $\mathbf{P}'_j$ are the deformed positions, $\mathbf{R}_i$ and $\mathbf{R}_j$ are the rotation matrices, $\alpha$ is a trade-off parameter for the regularization term, and $A$ is the surface area.

\section{Experiments and Analysis}\label{Experiments and Analysis}


In this section, we test our proposed method on three complex TVMs sequences used by MPEG: Levi, Dancer, and Basketball player, available in \cite{yang2023tdmd}. We also analyze a simpler TVMs sequence named Collision from \cite{MaxAffinityGlobalTracking}, which illustrates the collision and separation between a cuboid and a sphere, showcasing that our approach is not limited to human subjects. For our experiments, we use the first 30 frames of each sequence. To decrease the distortion and artifacts caused during the mesh alignment step as the number of meshes increases, we define every five frames as a subgroup, generating a reference mesh for each subgroup for the following evaluation and further processing. 

When applying ARAP volume tracking for all sequences, we set the $process$ mode and volume grid resolution to 512. Additionally, we use $eps = 1e-20$ for MDS to generate reliable sets of reference centers. For the RBFs, we use the $thin\_plate\_spline$ kernel, which, while not the optimal choice, interpolates points in 3-dimensional space relatively smoothly. However, the ideal mapping would be capable of handling discontinuities, and we are currently exploring such options.



We set $voxel\_size=0.1$ for the Basketball player sequence and $voxel\_size=0.01$ for the others when we use volume sampling to optimize the IoU to get the best aligned reference mesh. A $voxel\_size$ that is too large will result in overly coarse volume sampling, while a $voxel\_size$ that is too small will lead to significant computational costs.

We use the Hausdorff distance to evaluate the fitting errors of reference meshes. The experiments shown in Fig. \ref{Hausdorff distance evaluation} are done with the 0th to 4th frames for the Levi sequence, the 6th to 10th frames for the Dancer sequence, and the 15th to 19th frames for the Basketball player sequence. These specific meshes are showcased because they contain very close parts or self-contact regions. For complex TVMs sequences, using 500 centers is inefficient for accurately representing thin parts such as hands or calves. Therefore, 500 centers bring the most fitting errors and visual distortions. However, when the number of centers exceeds 1000, the mesh can be accurately represented using volume centers, leading to a smaller fitting error in the reference mesh. Additionally, the addition of GoF will increase the reference meshes' fitting error. Our method currently has difficulty in obtaining a visually satisfactory reference mesh when the GoF is greater than or equal to 5, whereas a GoF of 3 or lower yields acceptable results.

\begin{figure}[t]
    \centering
    \begin{subfigure}[b]{0.22\textwidth}
        \centering
        \includegraphics[width=\textwidth]{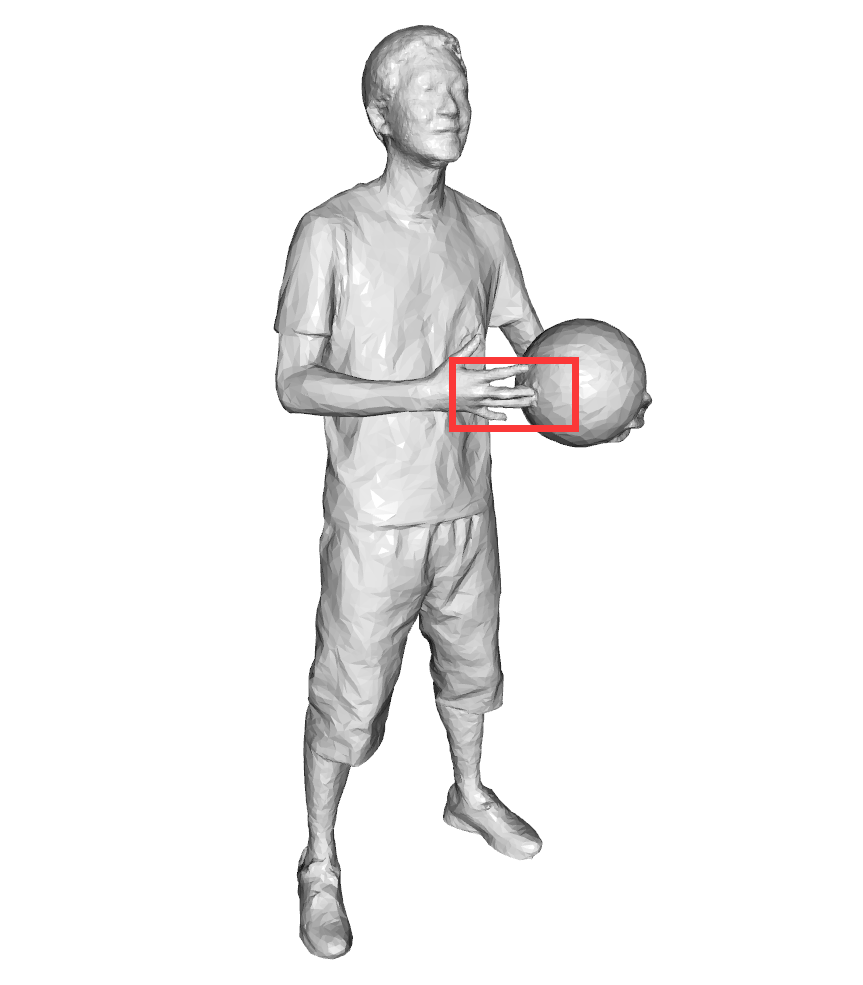}
        \caption{Original mesh of Basketball}
        \label{Visualoriginalbasketball}
    \end{subfigure}
    \hfill
    \begin{subfigure}[b]{0.24\textwidth}
        \centering
        \includegraphics[width=\textwidth]{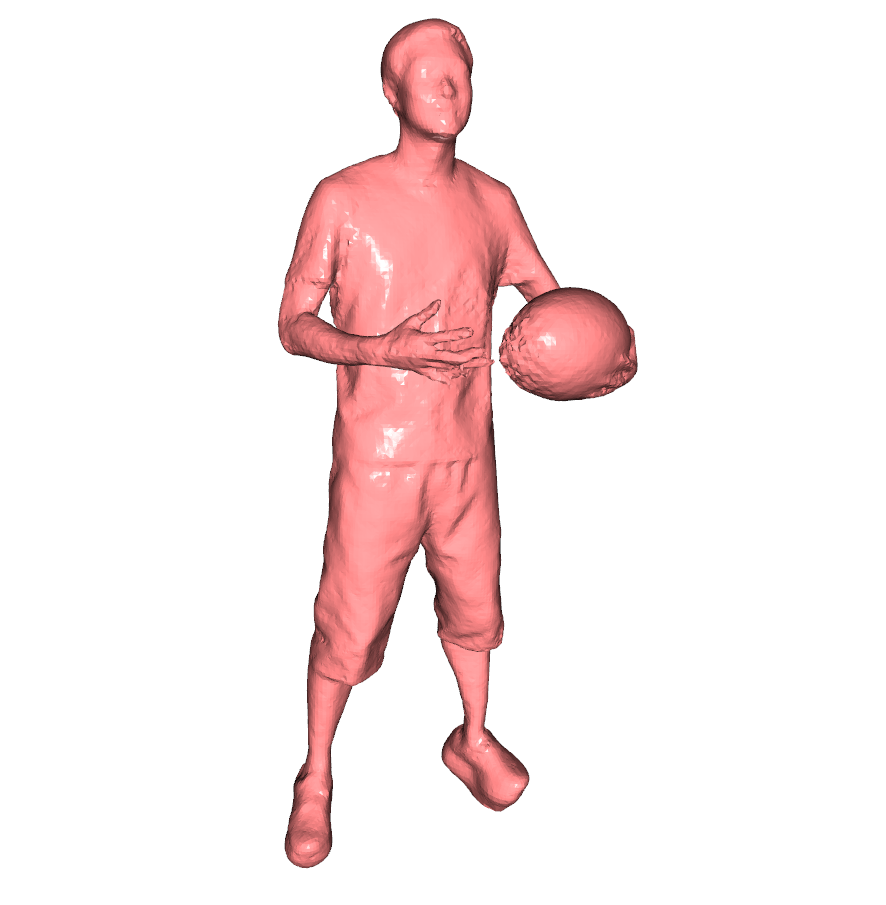}
        \caption{Reference mesh of Basketball}
        \label{VisualBasketball player}
    \end{subfigure}
    \hfill
    \begin{subfigure}[b]{0.22\textwidth}
        \centering
        \includegraphics[width=\textwidth]{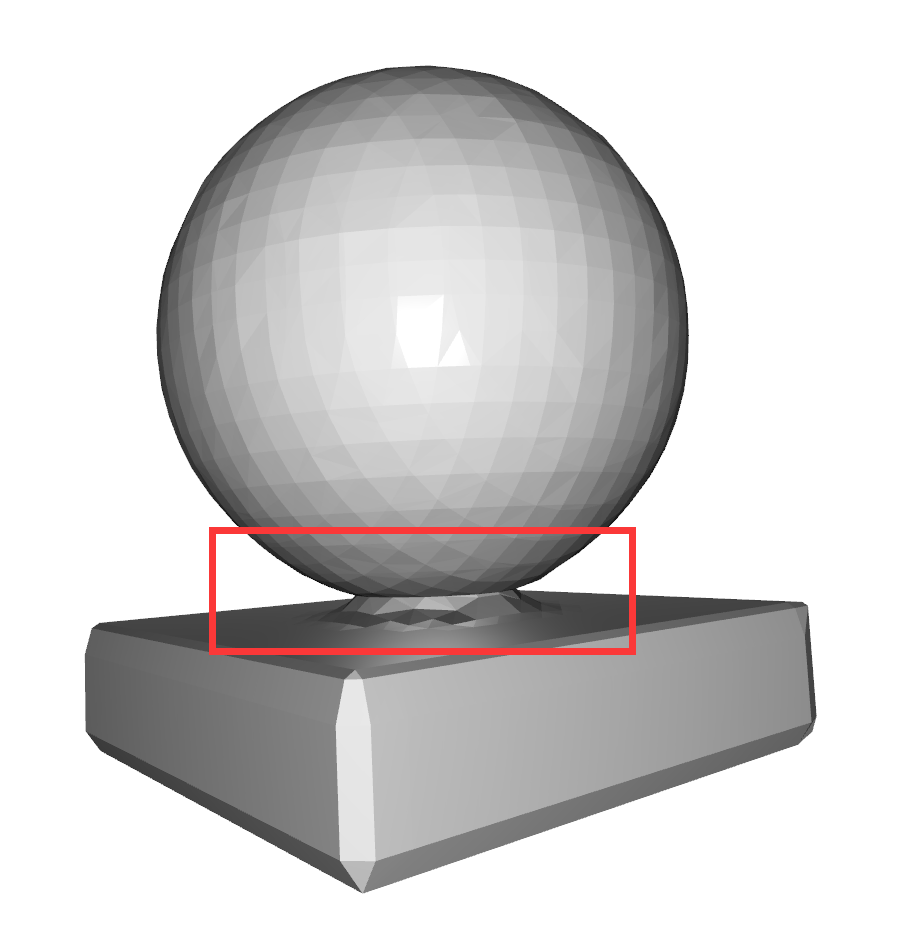}
        \caption{Original mesh of Collision}
        \label{Visualoriginalcollision}
    \end{subfigure}
    \hfill
    \begin{subfigure}[b]{0.24\textwidth}
        \centering
        \includegraphics[width=\textwidth]{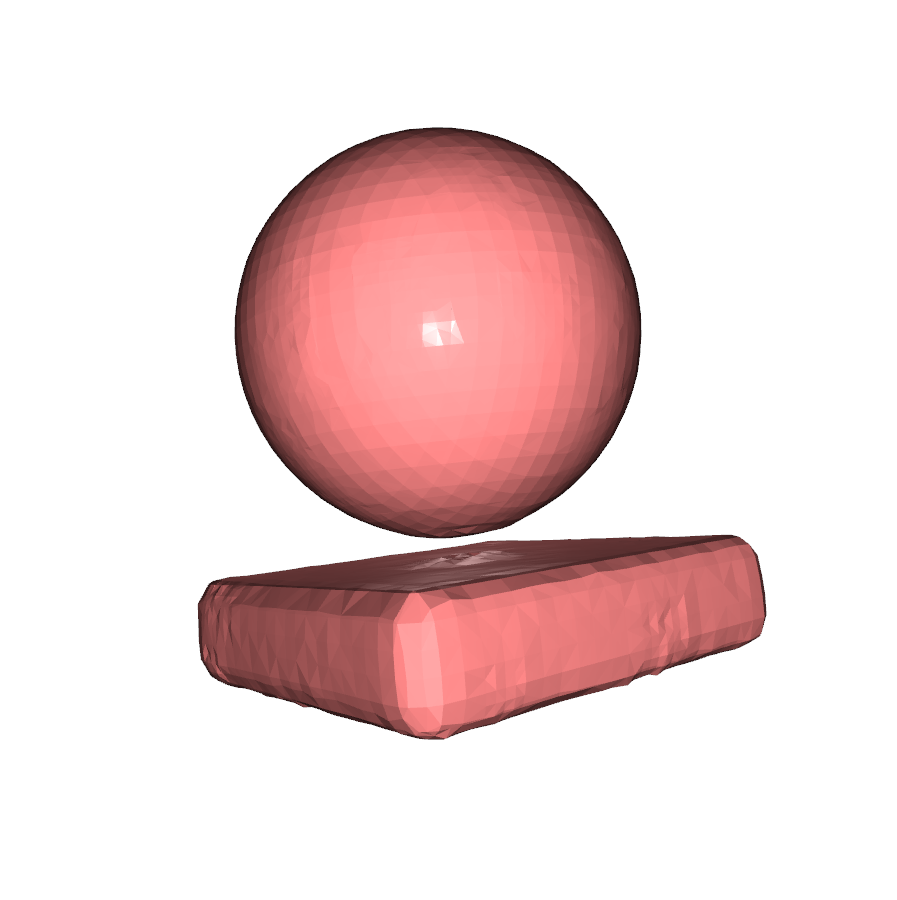}
        \caption{Reference mesh of Collision}
        \label{VisualCollision}
    \end{subfigure}
    \caption{Visual results of reference meshes. (a) and (c) show the original meshes in the Basketball player and Collision sequences, respectively, with self-contact regions marked by red boxes. (b) and (d) present the generated reference meshes for these sequences.}
    \label{Visual results}
    \vspace{-0.2in}
\end{figure}

Fig. \ref{Visual results} presents the original mesh and generated reference mesh for the Basketball player and Collision sequences. In Fig. \ref{Visualoriginalbasketball} and \ref{Visualoriginalcollision}, the original meshes include self-contact regions, while the reference meshes shown in Fig. \ref{VisualBasketball player} and \ref{VisualCollision} do not have self-contact regions. The red boxes highlight areas where the mesh has self-contact regions, such as where the hand touches the ball in (a) and where the sphere and the cuboid interact. Our method is able to separate these self-contact regions and get high quality reference meshes.

\begin{figure}[t]
\centerline{\includegraphics[width=0.35\textwidth]{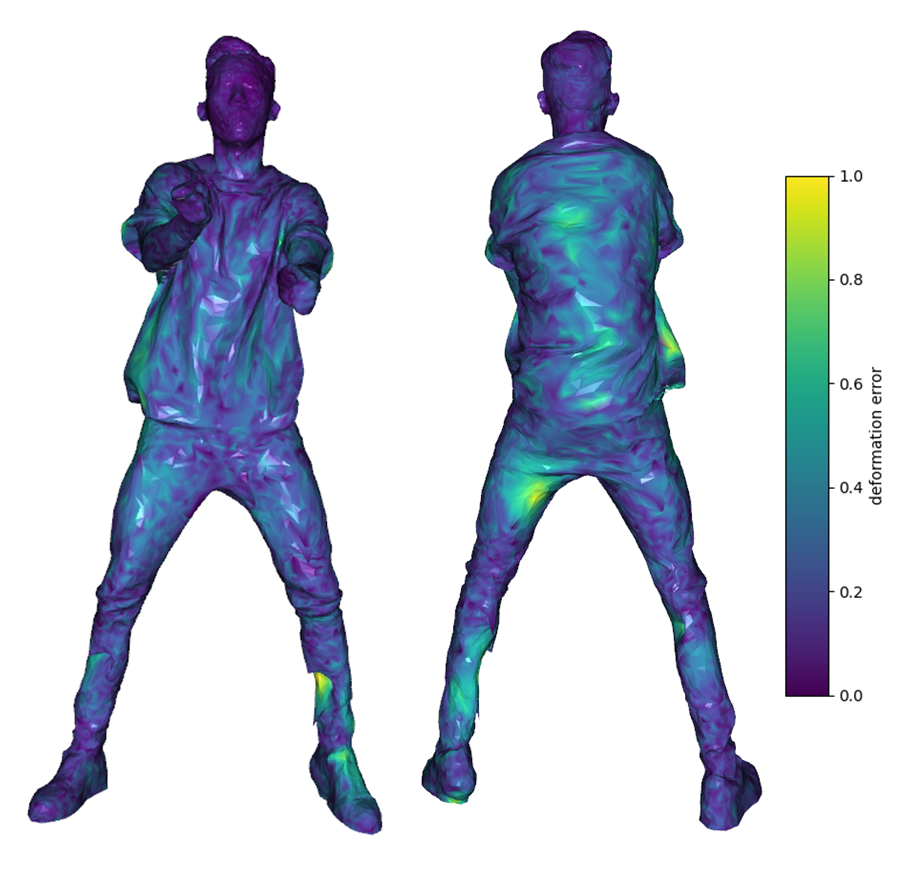}}
\caption{Visual evaluation of a deformed reference mesh compared to its target mesh using the Hausdorff distance metric. The color map, ranging from purple (low error) to yellow (high error), indicates the degree of deviation between the deformed mesh and its target.}
\label{deformationerror}
\vspace{-0.2in}
\end{figure}

Finally, we present the deformation results of the reference mesh as discussed in Section \ref{ARAPdeformation}, as evaluated by the Hausdorff distance. Fig. \ref{deformationerror} displays two views of the mesh: the front view and the back view. Areas with higher deviation are marked in yellow, while regions with minimal deviation are shown in purple. Fig. \ref{deformationerror} illustrates that our reference mesh can be deformed into one of the meshes within the group, with acceptable levels of distortion and deformation error.

\section{Conclusion}\label{Conclusion}
In this paper, we proposed a novel idea to extract a reference mesh without self-contact regions for TVMs, which can serve as the basis of many mesh processing methods that require reference mesh or need to perform deformation. We utilize volume centers that represent the surrounding areas to generate a tightly aligned reference mesh, however, the level of mesh alignment still needs further improvement. The mapping method RBFs can also be improved. It represents a smooth vector field, but we wish to represent mapping that tears apart components that are in self-contact. Our future research will further improve the quality of extracted reference mesh and apply it to various applications like mesh compression.

\vspace{12pt}
\color{red}

\end{document}